\documentstyle{mn}
\begin{document}


\title[Observations of HDF with ISO - I]{Observations of the
{\it Hubble Deep Field} with the {\it Infrared Space Observatory} -
I. Data reduction, maps and sky coverage
}
\author[Serjeant, Eaton, Oliver, {\it et al.}]
{Stephen Serjeant,$^1$ N. Eaton,$^1$ Seb Oliver,$^1$
A. Efstathiou,$^1$ Pippa Goldschmidt,$^1$
\vspace*{0.2cm}\\{\LARGE  
R.G. Mann,$^1$ B. Mobasher,$^1$
Michael Rowan-Robinson,$^1$
Tim Sumner,$^1$ 
L. Danese,$^2$
}\vspace*{0.25cm}\\ {\LARGE 
D. Elbaz,$^3$ 
A. Franceschini,$^4$
E. Egami,$^5$ 
M. Kontizas,$^6$ 
A. Lawrence,$^7$ 
R. McMahon,$^8$ 
}\vspace*{0.2cm}\\{\LARGE 
H.U. Norgaard-Nielsen,$^9$ 
I. Perez-Fournon,$^{10}$ 
I. Gonzalez-Serrano$^{11}$
}\\
$^1$Astrophysics Group, Imperial College London, Blackett Laboratory,
Prince Consort Road, London SW7 2BZ;\\ 
$^2$SISSA, Via Beirut 2-4, Trieste, Italy;\\
$^3$Service d'Astrophysique, Saclay, 91191, Gif-sur-Yvette, Cedex,
France;\\ 
$^4$Osservatorio Astronomico de Padova, Vicolo dell'Osservatorio 5,
I-35 122, Padova, Italy;\\
$^5$Max-Planck-Institut f\"ur Extraterrestrische Physik,
Giessenbachstrasse, D-8046, Garching bei Munchen, Germany;\\
$^6$Astronomical Institute, National Observatory of Athens, P.O.Box
200048, GR-118 10, Athens, Greece;\\
$^7$Institute for Astronomy, University of Edinburgh, Blackford Hill,
Edinburgh, EH9 3HJ;\\
$^8$Institute of Astronomy, The Observatories, Madingley Road,
Cambridge, CB3 0HA;\\
$^9$Danish Space Research Institute, Gl. Lundtoftevej 7, DK-2800
Lyngby, Copenhagen, Denmark;\\
$^{10}$Instituto Astronomico de Canarias, Via Lactea, E-38200 La
Laguna, Tenerife, Canary Islands, Spain;\\
$^{11}$Instituto de Fisica de Cantabria, Santander, Spain\\
}
\date{Accepted 1997 May 9;
      Received 1997 March 24;
      in original form 1996 Decmber 5}
\maketitle
\begin{abstract}
We present deep imaging at $6.7\mu$m and
$15\mu$m from the CAM instrument on the {\it Infrared Space
Observatory} ({\it ISO}), centred on the {\it Hubble Deep Field} ({\it
HDF}). These are the
deepest integrations published to date at these wavelengths in any region
of sky. We discuss the observation strategy and the data reduction. 
The observed source density appears to approach the CAM
confusion limit at $15\mu$m, and fluctuations in the $6.7\mu$m sky
background may be identifiable with similar spatial fluctuations in
the HDF galaxy counts. {\it ISO} appears to be detecting  
comparable field galaxy populations to the HDF, and our data yields
strong evidence that future IR missions (such as {\it SIRTF},
{\it FIRST} and {\it WIRE}) as well as 
SCUBA and millimetre arrays will
easily detect field galaxies out to comparably high redshifts. 
\end{abstract}


\section{Introduction} 

The {\it Infrared Space Observatory} ({\it ISO}, Kessler {\it et al.}
1996) offers 
an improvement of orders of magnitude in 
sensitivity over the IRAS satellite, at least at shorter wavelengths
($<20\mu$m). 
ISO is expected both to sample the  
intermediate redshift ($z\sim0.5-1$) star forming galaxy population at
lower, less extreme luminosities, and to detect more strongly star
forming galaxies to far higher redshifts. 
We used the CAM instrument on {\it ISO} (Cesarsky {\it et al.} 1996)
in Director's 
discretionary time to observe the 
{\it Hubble Deep Field} (HDF, Williams {\it et al.} 1996), resulting in 
the deepest surveys to date at $6.7\mu$m and $15\mu$m.
The obvious advantage of this field is the 
extensive multi-wavelength follow-ups either
published or underway 
({\it e.g.} Dickinson {\it et al.} 1997 in preparation, 
Fomalont {\it et al.} 1996, Cowie {\it et al.} 1996) as well as 
unparallelled deep multicolour optical morphologies from the {\it
Hubble Space Telescope} ({\it HST}).

The HDF galaxy population is strikingly dominated by blue objects
which may be comparable to local giant H{\sc ii} regions, and which
are consistent with 
significant star formation. This is re-enforced by their largely
disturbed structures. In such an interpretation a substantial
fraction of the luminosity from young massive 
stars is absorbed by dust and re-radiated in the
mid- and far-infrared. Clearly, imaging in the mid-infrared 
samples the spectral energy distributions of
such galaxies much closer to this significant, and perhaps dominant,
contribution to the bolometric power output. 

In this paper we present our CAM images of the HDF,
discuss the data reduction steps
and make crude comparisons of the sky fluctuations with 
smoothed HDF images. Subsequent papers will
address the source identification algorithm (paper II, Goldschmidt
{\it et al.} 1997); the source
counts and comparison with models (paper III, Oliver {\it et al.}
1997), including a more 
sophisticated treatment of the confusion noise; the associations with
HDF galaxies (paper IV, Mann {\it et al.} 1997); and the spectral
energy distributions of our 
sources and implications for star 
formation history (paper V, Rowan-Robinson {\it et al.} 1997). 

The maps presented here are available at {\tt
http://artemis.ph.ic.ac.uk/hdf} or from the authors. 

\begin{table}
\begin{tabular}{lll}
\hline
{\bf Parameter} & {\bf LW-2 $6.7\mu$m} & {\bf LW-3 $15\mu$m} \\ 
                    &              &       \\
Pixel field of view & $3''$        & $6''$ \\
M, N steps          & $8$, $8$     & $8$, $8$ \\
M, N step size      & $5''$, $5''$ & $9''$, $9''$ \\
$T_{\rm int}$       & $10$ sec     & $5$ sec \\
$N_{\rm stab}$      & 80           & 100 \\
$N_{\rm obs}$       & 10           & 20 \\
\hline
\end{tabular}
\caption{Summary of our CAM01 Astronomical Observation Templates
(parameters defining our $M\times N$ rasters). An exposure of 
$T_{\rm int}$ is made $N_{\rm obs}$ times at each of the $M\times N$
raster positions. Each raster was preceeded by $N_{\rm stab}$ readouts
at the first raster position to stabilise the CAM detector.}
\end{table}

\begin{table*}
\begin{tabular}{lllllllllll}
\hline
{\bf Band} & $\lambda$& {\bf Width} & {\bf Area} & {\bf Date}  & {\bf
$I_{\rm MODEL}$} & {\bf $I_{\rm ISO}$} & 
{\bf $1\sigma/$} & ~~~{\bf Normalisation} \\ 

            & ($\mu$m) & ($\mu$m) &    & {\bf Observed} & & & {\bf beam} &
{\bf Shift/Add~~Drizzle}\\
            &            &      &            &           &        & &       &           \\

LW-2        & $6.75$ &$3.375$ & WF-2 &28/6/1996& 73.4 &
77.3 & 7.78 & ~~1.03431~~~~~1.03390  \\

            &        &        & WF-3 &28/6/1996& &
79.9 & 8.59 & ~~1.01630~~~~~1.01492  \\

            &        &        & WF-4 &28/6/1996& &
80.0 & 8.43 & ~~1.00000~~~~~1.00000  \\

LW-3        & $14.5$ & $4.833$& WF-2 &1/7/1996 & 508 &
401  & 36.8 & ~~1.00000~~~~~1.00308  \\

            &        &        & WF-3 &1/7/1996 & &
402  & 27.4 & ~~1.00255~~~~~1.00000  \\

            &        &        & WF-4 &26/6/1996& &
396  & 46.1 & ~~1.01306~~~~~1.01454  \\
\hline
\end{tabular}

\caption{Summary of Calibration and Sky Backgrounds. Intensities
quoted are 
in $\mu$Jy per arcsec$^2$. The sky
backgrounds in each of the fields ($I_{\rm ISO}$) are compared against
the model ($I_{\rm MODEL}$) discussed in the text.
Noise estimates assume
a circular beam size of $6''$ for LW-2 and $12''$ for LW-3.}
\end{table*}

\section{Data acquisition and analysis}
\subsection{Observation Strategy}
The observing strategy was designed to reach the predicted confusion
limit of CAM, using the source count models of Pearson \&
Rowan-Robinson (1996). 
The observations were performed
in two bands, $6.7\mu$m and $15\mu$m (the LW-2 and LW-3 filters
respectively), to obtain (albeit limited) colour information. 
These bands are the widest available in
CAM so yield the deepest possible integrations. 
Typically more than one HDF galaxy falls within
the Airy disk, even at $6.7\mu$m, so sub-pixel offsets were
needed to maximise the available spatial information.

We made three $8\times8$ rasters in microscanning mode (CAM01)
analagous to the ``dithering'' in the HDF, each
centred 
on one of the HDF Wide Field (WF) frames, at both $6.7\mu$m and
$15\mu$m, using in 
total $\sim44.9$ks of time including overheads. Our choice of 
$3''$ ($6''$) pixel sizes at $6.7\mu$m ($15\mu$m) and raster
step sizes of $5''$ ($9''$) yields optimal flat fielding accuracy
and spatial resolution. At $6.7\mu$m this gives a $96''$ field of view. 
well-matched the HDF WF frames. 
The spacecraft jitter observed in-flight is $\pm0.5''$ ($2\sigma$
limits, halfcone), much 
smaller than our choices of either pixel size or step size. 
Cosmic ray transients (discussed below) ruled out
readout integration times longer than $10$ seconds, and the
signal-to-noise {\it vs.} time predictions from the CAM simulator
implied diminishing returns for more than $10$ ($20$) readouts per
raster position at $6.7\mu$m ($15\mu$m). The number of stabilisation
readouts prior to the rasters at each wavelength is appropriate for
$5\sigma$ sources. 
The Astronomical Observation Template (AOT)
parameters are summarised in Table 1; for a more detailed discussion
of these parameters see the CAM Observers Manual\footnote{The
CAM manual is available from
{\tt http://isowww.estec.esa.nl/manuals/iso\_cam/}}. 
Resulting noise levels in each of the fields are listed in table 2. 
The pixel scale of ISOPHOT made longer wavelength observations
impracticable. 


The edited raw data FITS files (supplied by
ESA) were processed using the
CIA (CAM Interactive Analysis April 1996 version) IDL package with
the exception of the deglitching, the construction of the flat
field and the mosaicing of the rasters, as discussed below.

\subsection{Dark subtraction, deglitching and flat fielding}
The default dark frame in the April 1996 CIA version was subtracted
from the data. 

Cosmic ray events were easily identified in the readout histories of
each pixel as $>4\sigma$ rises followed (one or two readouts
later) by $>4\sigma$ falls. A similar algorithm was used to find
readout troughs. These events were masked out in the
mosaicing discussed below.

However, a minority of cosmic
rays appear to cause transients in subsequent readouts with 
roughly exponential decays (see figure 1) persisting over a few
readouts. These glitch transients are in general difficult to model.
No attempt was
made to identify and remove them; instead, they are effectively
removed by median filtering in the mosaicing below. 

Use of the ESA-supplied flat field gave very unsatisfactory
results. Instead, we created our own sky flat by noting that each
detector pixel samples $64$ different sky positions during the
raster. For each detector pixel, we examined the histogram of
(unmasked) readouts and fitted Gaussians to find a mean value. To
eliminate both sources and glitch transients, $>5\sigma$ outliers from
the mean were eliminated and the fit was iterated.
 

\subsection{Shift-and-add maps}

The rasters were mosaicked together using two competing algorithms:
drizzling (described below), and variants on shift-and-add. The latter
are expected to 
have higher signal-to-noise but at the expense of
spatial resolution. 
 
In order to eliminate possible long timescale detector sensitivity
drifts, the rasters were renormalised by the following method. First,
we calculated the median readout in each of the $32\times32$ CAM
pixels for each raster at each wavelength. Second, we found the mean
in the central $11\times11$ of these images. This determined the
relative renormalisations, which are listed in Table 2.
Comparison of the source positions 
in each individual raster revealed a
systematic offset in the WF-4 $15\mu$m frame, probably due to a random
offset in the lens positioning. For consistency with the drizzled
mosaics we applied the same offset to this raster as discussed below
for the drizzling. 

In the shift-and-added frames, we began by defining an image with pixel
size one-sixth that of the CAM detector pixels (ie, one-sixth of $3''$ at
$6.7\mu$m, and of $6''$ at $15\mu$m), encompassing the area surveyed
by all three rasters. Each position in this fine-gridded image may
have been observed several times by the CAM detector array, so we
compiled a list of such CAM pixel readouts for each fine-gridded image
position. The data at a given sky position could then (for
example) be median filtered or mean averaged to produce a final image. 
 
The glitch transients discussed above 
make a large contribution to the noise in final mosaics 
made by simple mean averages of readouts. An obvious alternative is median
filtering, though this has a signal-to-noise penalty (about
$\sqrt{2}$). The 
glitch transients have timescales of order or less than the duration of a
pointing, so a possible compromise is to mean average over pointings, and
median filter the means. This may also have the advantage of identifying
affected readouts more efficiently. 

Several mosaics were therefore created from the readout arrays: (a) using
the median readout at each position, to eliminate glitches; (b) using the
mean readout at each position, with the $\pm5\sigma$ outliers eliminated
and iterated to eliminate glitches; (c) using the mean readout within each
raster position, followed by the median of these means; (d) using a
clipped, iterated mean of pointing means. We found the best $6.7\mu$m
map to be 
the simple median filtered image, but at $15\mu$m the optimal map is the
median-of-means, presumably reflecting a greater sensitivity to glitch
transients in the final mosaic.

\subsection{Drizzled maps}

The drizzled images were produced with the same code (Fruchter \&
Hook, 1996)
used to produce the optical {\it HST} images. Briefly, instead of
superimposing overlapping pixels, the drizzle algorithm allows the
user to shrink the input pixel sizes (the ``footprint'') before
superimposing. At one 
limit the drizzle algorithm is equivalent to interlacing; at the other
it is similar to shift-and-add above.
Since the {\it ISO} HDF images were
taken with fractional pixel spacings between them maps with increased
resolution can be constructed. The footprint was chosen to produce
roughly the
highest resolution image without leaving gaps in the output image. 


Glitch transients were first removed from the input images using the
following median filtering scheme, noting that at
$6.7\mu$m the raster step offsets correspond to $1\frac{2}{3}$ pixels
and at $15\mu$m the offsets are $1\frac{1}{2}$ pixels. The 64 individual
pointings were grouped into sets whose members have
integer pixel offsets
(9 groups for LW-2 and 4 for LW-3), and for each of these groups the medians
of the data at each pixel position (allowing for the shifts) was put into an
image and the variance was computed. These 9 images at $6.7\mu$m and 4 images
at $15\mu$m served as input to the drizzling algorithm. 

The satellite astrometry (RA, Dec and spacecraft roll angle) from the
individual pointings 
was combined into these (9+4) data sets by taking the average of those
pointings making up each group, after shifting them by the nominal raster
offsets. This astrometry was used to define the tangent-plane projection
used in the final maps.

The {\it ISO} images were drizzled assuming there were no geometrical
distortions 
in the images so that rows and columns were always assumed to be parallel
and equally spaced. The $6.7\mu$m images were sub-sampled to $1/3$ of the
input pixel size, resulting in output pixels of $1''$. The input pixel
footprint was set to $0.65$ and the pixel weights were taken from the
reciprocals of the variances. The $15\mu$m images were sub-sampled to $1/2$
of the input pixel size, resulting in output pixels of $3''$. The
footprint was again set to $0.65$ and the same weighting scheme was used.

The drizzle code preserves flux by sharing the input levels amongst the
output pixels. To return the output images to intensity units the
$15\mu$m images were multiplied by $4$ and the $6.7\mu$m images by $9$.
The input weights were the reciprocals of the variances, and to convert the
output weights from the drizzle code back into a variance a multiplying
factor of $1/$footprint$^2$ ($2.367$) had to be applied. These output
variances are not totally independent as each has contributions from
neighbouring pixels.
 
The $15\mu$m (LW-3) image for field WF-4 appears to have incorrect
astrometry in 
that when the appropriate offset was applied the sources did not line up
exactly with the sources in field WF-2 or WF-3. An empirical offset was
calculated using the {\tt correl\_images} routine from the IDL astrolib
library ({\tt http://idlastro.gsfc.nasa.gov/homepage.html}) to
cross-correlate the images and from visual checking. The resultant shift
was $6.4''$ in the detector x-direction and $-0.7''$ in the
y-direction.
 
To ensure the final mosaics were not affected by global variations in
the background the medians from the drizzled images of the individual
pointings were compared and a multiplying factor (listed in Table 2)
was applied to all the 
frames making up a pointing to bring them to the same level. The final
drizzled images were then recomputed.

\subsection{Flux calibration}
To estimate the sky background, we constructed
histograms of pixel counts for the central third of the unnormalised
drizzled maps 
of each individual field. We fitted Gaussians to each of these
histograms, and the resulting sky levels are listed in Table 
2. The relative sky levels are in excellent agreement with the
relative normalisations calculated above. A slightly more
sophisticated procedure was adopted for the noise estimates. We
selected a grid of positions in the central $\sim1/3$ of each drizzled
mosaic, and placed a circular aperture of the Airy disk size on each
position. We could then estimate the noise level on the scale of the
Airy disk from the histogram of counts enclosed by these apertures. 

Flux calibration assumes the standard conversion in the CAM
handbook. In Table 2 we compare our background measurements with
a zodiacal background
model. This model linearly interpolates between entries in Table 8 
of the CAM Observer's Manual, but assumes a $275$K blackbody 
zodiacal spectrum (Hauser {\it et al.} 1984) rather than the
alternative in the manual. The 
solar elongation angle at the time of our observations
was approximately $71.3^\circ$. The cirrus contribution at both
wavelengths is expected to be $<2$ per cent of the zodiacal
background. 

The sky calibration appears on the whole to be quite satisfactory. 
Checks on the calibrations using the sky, such as these, do
not of course check the linearity of the response to sources, 
which may be very difficult to address with the {\it ISO} HDF data. 
It may nevertheless be possible to estimate this by 
examining the average temporal profiles of detected sources. 
Further discussion on the sensitivity to point sources is contained in
paper II (Goldschmidt {\it et al.} 1997). 

\section{Results}

\subsection{Comparison of maps}
Figures 2 and 3 show the results of the drizzling
algorithm. The correspondence between the drizzling and the
shift-and-add mosaics (not shown) is excellent suggesting that neither
algorithm is introducing artifacts. Figures 4 and 5 show the coverage
maps at the two wavelengths. Note the poorer coverage at the edges,
which causes the poorer signal-to-noise at the corresponding edges of
the mosaics. 

\subsection{Source confusion}
Several sources are clearly seen at $15\mu$m, and the source
identification will be discussed in a later paper. However, we note
here that a visual inspection 
clearly shows that we are approaching the confusion limit of one
source per $40$ beam sizes.

One surprising feature in the $6.7\mu$m maps is the apparent sky
fluctuations at around $3\%$ of the background,
which may be caused by source confusion, or by glitch transients which
have been smeared out by the mosaicing process. We tested this by taking
ratios of each of the $6.7\mu$m shift-and-add mosaics. One apparent
(albeit dubious) source 
was less prominent with increasing median filtering, so is a good
candidate for such a smeared-out transient. 
Apart from this the ratios of the $6.7\mu$m shift-and-add
mosaics are flat to $\sim1\%$ accuracy,
and do not have structures corresponding even 
roughly to the observed $6.7\mu$m maps. This 
suggests that the fluctuations are not an artefact of transients or
flat-field errors, but are due to marginally detected sources. 

A more convincing demonstration was made by shuffling the CAM pixels at
random and making mosaics. The resulting frames should be equally
susceptible to smeared-out transients, but genuine sources and background
structure should be dispersed evenly over the mosaic. These randomised
frames show much less structure than the real frames ({\it e.g.},
figure 6), again suggesting that
this structure is not an instrumental artefact. 
 
Note however that
this does not test for the presence of correlated pixel-to-pixel flat
field variations. If such transient effects are present, they would be
indistinguishable from genuine sky structure in the mosaicked images,
and would be extremely difficult to remove in the data reduction
process without also removing genuine structure. (It is nevertheless
not clear how such correlated fluctuations would arise.) One approach
to testing this is to rotate or reflect the arrangement of the CAM
detector pixels, but repeat the same mosaicing as before. Any
sky structure will be dispersed over the image (though not necessarily
evenly), but the resulting images would be equally sensitive to
structure from correlated 
pixel-to-pixel fluctuations. These images should therefore show less
structure than the correctly mosaicked images. 
Figure 6 shows an example of
this test, and the frame does indeed show less structure than the
correctly mosaicked frame. However, it is not clear how
much of the remaining structure is due to unevenly dispersed, genuine
sky structure, and how much is due to correlated pixel noise. These
potential artefacts are still under investigation, but in the meantime
it should be noted that some of the structures may potentially be
due to instrumental artefacts.

Finally, a comparison of suitably smoothed drizzled and {\it HST} F814W images,
discussed in the next section, shows remarkable correspondence. 

\subsection{Field distortion}
To compare the apparent low level sky fluctuations with the
underlying HDF galaxy distribution, we convolved the drizzled {\it ISO}
mosaics with an empirical point spread function (Oliver {\it et al.}
1997, Goldschmidt {\it et al.} 1997). 
In figures 7 and 8 we show a greyscale reproduction of the F814W HDF
and flanking fields\footnote{The flanking field image (unfortunately
without astrometry) is obtainable from\\
{http://www.stsci.edu/ftp/science/hdf/project/flanking.html}}. 
Contours of our smoothed drizzled {\it ISO} mosaics are overlaid. 
Two important points should be noted: first, the apparent
low level structures do indeed match the inhomogenous galaxy
distribution in the {\it HST} F814W frame, at least at $15\mu$m; second,
this match is not perfect, 
suggesting an unknown astrometric error. 

The astrometric errors do not appear to be random, but rather
systematic and varying continuously over the field. 
This is unlikely to be due to telescope pointing and lens positioning
errors, since these would produce a non-varying
systematic offset (recall that the individual rasters were registered
with respect to each other). 
A likely explanation for the slight astrometric errors is therefore 
field distortion in CAM, {\it i.e.} the CAM detector array is not
quite square. In-flight calibration of this distortion by the CAM
consortium is currently underway; preliminary results indeed suggest
that the total field distortion is about one pixel over the whole
array, and that the distortion is continuous.
Once this distortion is quantified we expect to release appropriately
revised mosaics. In the meantime, further papers in this series will
conservatively treat the astrometry as subject to random errors of
order the Airy disk size. 

At $6.7\mu$m the stucture, if
present, is close to the noise; nevertheless, there are clearly
several marginally detected sources at the positions of bright
galaxies in the HDF F814W image: for example, several bright
galaxies in 
field 2 (top left HDF frame) lie on or close to contour peaks.
Note that the signal-to-noise decreases
sharply towards the edges, resulting in many features without apparent
I-band counterparts. Finally, recall that correlated pixel-to-pixel
flat field 
variations (if present) would mimic real sky structure. Nevertheless, at
least some of the stucture appears to be marginally detected or unresolved
sources.

\section{Discussion}
Visual inspection of the maps at $6.7\mu$m and $15\mu$m clearly shows
many sources and these will be discussed further in the following papers.
At these wavelengths the radiation is increasingly dominated by the 
re-processing of starlight by dust and it is to be expected that the
detected galaxies have above average star-formation rates.
Our data demonstrates that {\it ISO} can detect galaxy populations comparable
to those in faint optical surveys, and may
also have implications for proposed confusion
limited surveys at around our wavelengths, such as that from the {\it
WIRE} satellite. However, we defer a more extensive discussion of the
confusion limit to paper III (Oliver {\it et al.} 1997).

These results are also promising for surveys at other wavelengths. 
Mobasher {\it et al.} (1996) combined their HDF photometric redshift
database with a starburst spectral energy distribution model and the
HDF galaxy counts, to obtain 
predictions for cumulative number counts of HDF galaxies at
wavelengths longward of $6.7\mu$m. 
Their expectation of 
$\sim$ tens of sources at {\it ISO} wavelengths at our approximate limiting
flux densities per beam is clearly seen to be broadly correct. 
A detailed comparison with source count models is included in
paper III (Oliver {\it et al.}
1997); however, 
we note here that Mobasher {\it et
al.} also used the 
same models to predict $\sim$ a hundred sources in the HDF at both
$60\mu$m and $0.8$mm to
$\sim10-100\mu$Jy levels. Our results are thus clearly encouraging for
SCUBA and the proposed large milimetre arrays, as well as for planned
future infrared space missions such as FIRST
and SIRTF. 

Further information on the {\it ISO} HDF project can be found on the
{\it ISO} HDF
WWW pages: {\tt http://artemis.ph.ic.ac.uk/hdf/}

\section*{Acknowledgements}
This paper is based on observations with ISO, an ESA project with
instruments funded 
by ESA member states (especially the PI countries: France, Germany,
the Netherlands and the Unitied Kingdom) and with participation of
ISAS and NASA.
This work was supported by PPARC (grant number GR/K98728) and by the
EC TMR Network programme (FMRX-CT96-0068). 

\begin{figure*}
\parbox{150mm}{
Figure 1:
Example glitches. The data shown is a dark subtracted pixel history
with the (logarithmic) y-axis in the instrumental analogue-to-digital
units per second 
(ADU/s) and x-axis in readouts. Note the clear glitches 
as well as the bright glitch at readout $\sim 580$ with a
transient. 
}
\end{figure*}

\begin{figure*}
\parbox{150mm}{
\hspace*{0cm}\parbox{120mm}{\vspace*{0cm}Figure 2: 
Drizzled mosaic at $6.7\mu$m. North is approximately 10
degrees right of vertical and east is to the left.
}}
\end{figure*}

\begin{figure*}
\parbox{150mm}{
\hspace*{0cm}\parbox{120mm}{\vspace*{0cm}Figure 3: 
Drizzled mosaic at $15\mu$m.
}}
\end{figure*}

\begin{figure*}
\parbox{150mm}{
{Figure 4:
Coverage map at $6.7\mu$m. Note the poor coverage at the edges. The
scale converts the grey levels to the relative coverage. 
}}
\end{figure*}
\begin{figure*}
\parbox{150mm}{
{Figure 5:
Coverage map at $15\mu$m. As with the $6.7\mu$m counterpart, the
coverage is poor at the edges.  
}}
\end{figure*}

\begin{figure*}
\parbox{150mm}{Figure 6: 
Comparison of the $6.7\mu$m drizzled image (left) with a
similar (shift-and-add) mosaic made after randomising the CAM detector
pixel positions 
(right), and a drizzled mosaic obtained after rotating the CAM
detector array through 
$90^\circ$ (centre). As argued in the text, the drizzling and
shift-and-add methods (not compared here) are found to be in excellent
agreement. Note the lack of structure 
in the randomised frame, and the less prominent structure in the
rotated-detector frame. As discussed in the text, the latter attempts
to disperse any true 
structure on the sky over the image, while being equally sensitive to
correlated pixel noise. 
}
\end{figure*}

\begin{figure*}
\parbox{150mm}{
{
Figure 7a: 
The {\it HST} F814W image overlayed with contours of the smoothed drizzled
$6.7\mu$m image discussed in the text. 
}}
\end{figure*}

\begin{figure*}
\centering
\parbox{150mm}{
Figure 7b: 
The smoothed drizzled
$6.7\mu$m image discussed in the text, with contours overlaid.
}
\end{figure*}

\begin{figure*}
\centering
\parbox{150mm}{
\vspace*{0cm}Figure 8: 
The {\it HST} F814W image overlayed with contours of the smoothed drizzled
$15\mu$m image discussed in the text. 
}
\end{figure*}

\label{lastpage}
\end{document}